\documentclass[intlimits,twoside,a4paper]{article}

\usepackage{graphicx}
\usepackage[cp1251]{inputenc}

\usepackage[eqsecnum]{cmpj3}
\issue{2022}{25}{3}{33201}
\doinumber{10.5488/CMP.25.33201}
\title[ MC simulations in boxes with variable shape]%
{ On the algorithm to perform Monte Carlo simulations in cells
with constant volume and variable shape }
\author[A. Baumketner]{A. Baumketner\orcid{0000-0003-2726-931X}\thanks{Corresponding author: \email{andrij@icmp.lviv.ua}.} } 
\address{
 Institute for Condensed Matter Physics of the National Academy of Sciences of Ukraine,
 1 Svientsitskii Str., 79011, Lviv, Ukraine 
}

\newcommand{\ds}{\displaystyle}
\date{Received May 02, 2022, in final form May 02, 2022}

\begin{document}

\maketitle

\begin{abstract}
In simulations of crystals, unlike liquids or gases, it may happen that
the properties of the studied system depend not only on the volume of the
simulation cell but
also on its shape. For such cases it is desirable to 
change the shape of the box on the fly in the course
of the simulation as it may not be known ahead of time which geometry
fits the studied system best. In this work we derive an algorithm
for this task
based on the condition that the distribution of specific geometrical
parameter observed in simulations at a constant volume matches that
observed in the constant-pressure ensemble. The proposed algorithm is
tested for the system of hard-core ellipses which makes lattices of 
different types depending on the asphericity parameter of the
particle. It is shown that the performance of the algorithm critically
depends on the range of the sampled geometrical parameter. If the range
is narrow, the impact of the sampling method is minimal. If the range is
large, inadequate sampling can lead to significant distortions of the
relevant distribution functions and, as a consequence, errors in the
estimates of free energy.

\keywords{ 
hard-ellipse fluid, Monte Carlo simulation, constant
volume, varying shape, umbrella sampling}
%
%\pacs 02.70.-c, 61.20.Ja, 61.20.Gy

\end{abstract}

\section{Introduction}

Today computer simulations play a key role in fundamental research across multiple disciplines, including
physics, chemistry and materials science~\cite{gunst}.
A large share of computational studies 
employ simulation boxes with fixed volume.
This choice is mainly motivated by convenience as
constant-volume/constant-temperature ensemble is easier to program than the
equivalent ensemble with constant
pressure. But this is also due to the involvement of the constant-volume ensemble
in other, specialized simulation techniques such as free energy calculations~\cite{r2},
Gibbs ensemble~\cite{gemc} or replica-exchange method~\cite{rex1,rex2,rex3}.
Regardless of the particular context, it is always understood that 
the effect of volume vanishes 
in the thermodynamic limit 
where the results are thought to be 
independent of the employed ensemble. 
This claim is certainly true for 
liquids or gases, whose 
properties are independent of the geometry of the box.

\begin{figure}[htbp] 
\centering 
\includegraphics[width=0.5\textwidth]{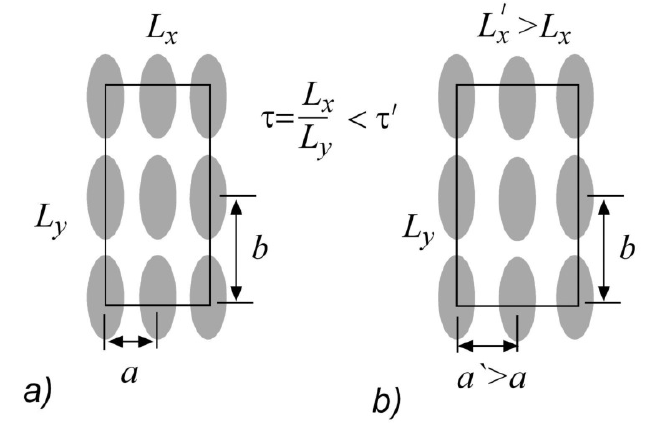} 
\caption{ 
Aspect ratio of simulation boxes is governed by the geometry of
the modelled lattice. Incorrect initial guess will result in lattice
distortions in constant volume simulations, for instance leading to
$a' > a$. To circumvent this problem, the
box should be capable of changing its shape.
} 
\label{f1} 
\end{figure}

In the case of crystals, however, the situation could be quite different~\cite{fortini,vega}. 
In crystalline materials there could be
properties that depend explicitly on the volume as well as on the shape
of the box. Take for instance the example of a rectangular lattice with
lattice constants $a$ and $b$, as shown in figure~\ref{f1}. The
dimensions of the box that accommodates $n$ columns and $m$
rows are $L_{x} = a n$ along $x$ axis and $L_{y} = b m$ along
$y$ axis as shown in figure~\ref{f1}~(a). The corresponding aspect ratio
is \(\tau = L_{x}/L_{y} = an/bm\). Now, let us
assume that the lattice constants are not known ahead of time but are
meant to be determined in the course of the simulations. If we initially
choose a box with the wrong aspect ratio, \(\tau' > \tau\) for
instance, see figure~\ref{f1}~(b) for appropriate illustration, the lattice
constant determined in simulations will also be incorrect. 
Since the geometry of the cell drives the structure of the lattice,
wrong geometry translates into wrong structure.
Importantly, lattice distortions
will not go away easily even when the size of the simulation
box is increased.

One way to deal with this issue is to estimate the free energy of the
studied system \(F\) as a function of some geometrical parameter of the
cell, for instance \(\tau\), and then choose the geometry with the
lowest \(F\). This can be done by a variety of tools, including the
Einstein crystal method for the free energy
computation~\cite{r1,r2}. Although formally correct, this
approach is cumbersome and carries a large computational cost. An
alternative is to allow the shape of the box to change in the course of
the simulation. The proper geometry then corresponds to the free energy
minimum, so it will be seen as the most frequently visited structure. An
additional benefit will be for systems that can populate multiple
geometries at the same time as their relative free energy in this case can be
determined from a single Monte Carlo (MC) trajectory.

The question then is how does one allow the shape of the box to change?
How is that accomplished in practice? Would, for instance, generating
randomly, from time to time, a new aspect ratio \(\tau\) in the course
of the simulation constitute a good method? If not, what is the good
method? 
These questions had been addressed before as multiple studies
report using simulation cells with constant volume but variable
shape~(MCVS)~\cite{frenkel,bolhuis,fortini,vega}.
Unfortunately, the details of the performed simulations are scarce and to
establish the specifics of the used algorithms appears difficult.
Yet, we find evidence that the method one employs to sample the trial geometries
may have measurable 
consequences for physical properties extracted
from simulations.
Thus, which geometry parameters are best to choose in MCVS simulations, 
and how to choose them remains
unclear.
These are the questions that we answer in the present article. We show
that in order for the constant-volume ensemble with variable shape to be
consistent with the constant-pressure ensemble, the aspect ratio should
be sampled from the $1/\tau$ distribution. Any other sampling law
will lead to erroneous results. We illustrate this point for the system
of impenetrable ellipses in two-dimensional space, for which we evaluate
the performance of the method that relies on uniformly sampled
$\tau$, or on the so-called \(\tau\)-sampling and show that it
produces wrong free energy for the relevant states of the system.

\section{Theory}
\label{sec:Theory}
\subsection{Sampling law for $\tau$ }

Let us assume that, in addition to volume, the partition function in the
canonical, or NVT for short, ensemble
$Q(N,T,V;\tau) = \int_{V; \tau} \rm{exp}[{- \beta U(\Gamma)}]\rd \Gamma$
explicitly depends on some geometrical parameter~$\tau$, for
instance the aspect ratio of the box sides. Here, \(\beta\) is the
inverse temperature, $V$ is the volume, \(U(\Gamma)\) is the
potential energy and \(\Gamma\) is the abbreviation for the point in
the configuration space. The integration is carried out over volume
$V$ with the set parameter $\tau$. The full partition function then
should be constructed as a weighted sum (or integral) over all possible
realizations of the additional degree of freedom~\cite{smit}. In the most general
case, one finds:
\begin{equation}
Q(N,T,V) = \int_{}^{}{Q(N,T,V;\tau)f(\tau)\rd \tau = \int_{}^{}{P(\tau) \rd \tau}},
\label{e1}
\end{equation}
where $f(\tau)$ is the weighting function of the extended ensemble
defined by both volume and the shape of the box and \(P(\tau)\) is the
probability distribution function of \(\tau\), characteristic of the
NVT ensemble with variable shape. In principle, one is free to
choose $f(\tau)$ arbitrarily, provided that it satisfies
certain conditions typically imposed on distribution functions, such as
positive definiteness or integrability. Here, we will select
$f(\tau)$ on the condition that the extended ensemble with
fixed volume satisfies distribution of \(\tau\) specific for the
constant pressure, or NPT, ensemble. In this way, modelling in the
two ensembles, constant-volume and constant-pressure, will be consistent,
hence minimizing finite size effects.

The distribution function in the constant-pressure ensemble reads:
\begin{equation}
P(\Gamma,N,T,{\mathcal P})\sim \re^{- \beta {\mathcal P} V}\ \re^{- \beta U(\Gamma)},
\label{e2}
\end{equation}
where $\mathcal{P}$ is the pressure. Let us focus now on 2D space and obtain
formulas for this simpler case first. The relevant volume is
\(V = L_{x}L_{y}\sin\alpha\), where \(L_{x}\) and \(L_{y}\) are the
lengths of the simulation box and \(\sin\alpha\) is the sine of the
angle between them. In NPT simulations, volume is sampled randomly
from a uniform distribution. This can be achieved either by uniformly sampling
one of the variables involved in volume, \(L_{x}\), \(L_{y}\)
or \(\sin\alpha\), or by non-uniformly sampling some combination of
these variables which leads to a uniformly distributed
volume~\cite{r2}. Let us assume that the first scenario takes
place, as it is more general and can be applied to both liquids and
crystals, and each concerned variable is sampled uniformly, one at a
time and in random order. The joint distribution function measured in
such simulations for variables \(L_{x}\), \(L_{y}\) and \(\sin\alpha\)
will be given by the following expression:
\begin{equation}
P\left( L_{x},L_{y},\sin\alpha \right)\sim \re^{- \beta {\mathcal P}L_{x}L_{y}\sin\alpha}\ Q(N,T,V;\tau),
\label{e3}
\end{equation}
which can also be obtained by integrating distribution function~(\ref{e2}) over
all configurations \(\Gamma\). The dependence on $\tau$ in the partition function arises because the
integration
over $\Gamma$ is carried out for the box with specific dimensions $L_x$ and
$L_y$, which
in addition to the volume $V=L_x L_y$ also define other geometrical
parameters
including $\tau = L_x/L_y$.

\begin{figure}[htbp] 
	\centering 
	\includegraphics[width=0.7\textwidth]{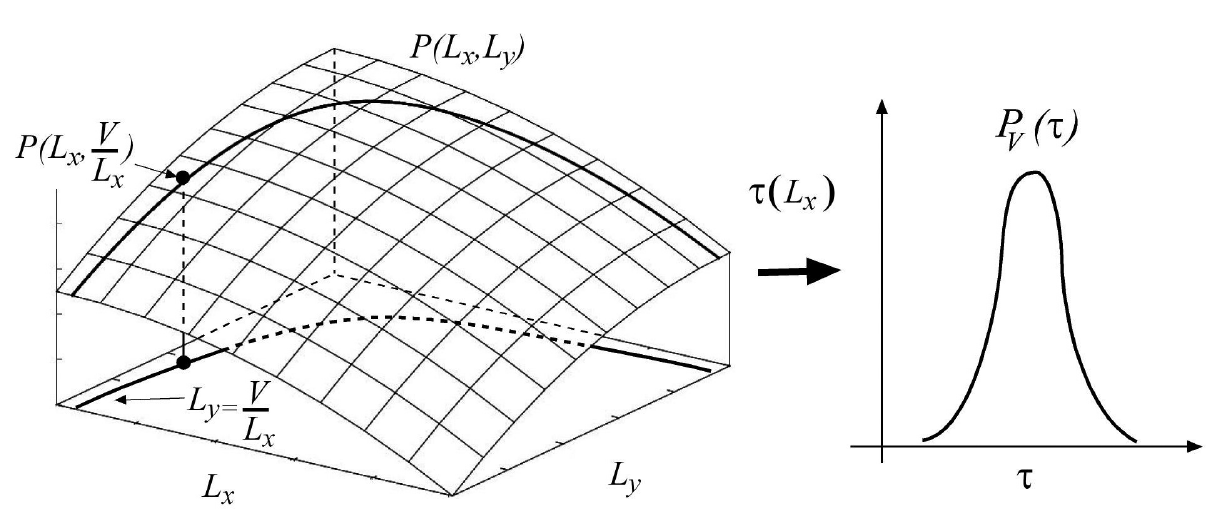} 
	\caption{ 
		Cartoon illustrating hypothetical joint distribution function of
		variables $L_x$ and $L_y$ 
		for the constant-pressure ensemble. Constant-volume
		configurations correspond to the sub-ensemble bound to the line
		$L_y=V/L_x$,
		where $L_x$ 
		is treated as the independent variable.
		Sampling along this line uniquely defines the distribution function
		$P_V(\tau)$ for 
		geometrical parameter $\tau$.
	} 
	\label{f2} 
\end{figure}

As an illustration, consider a schematic distribution
\({P(L}_{x},L_{y})\) shown in figure~\ref{f2}, specific for rectangular boxes
with \(\sin\alpha = 1\). Among all possible configurations, those that
correspond to volume $V$ satisfy the constraint \(V = L_{x}L_{y}\),
creating a one-dimensional sub-ensemble characterized by a single degree
of freedom. If \(L_{x}\) is chosen as the independent degree of freedom,
configurations with given $V$ and \(L_{x}\) will appear in
simulations with probability \({P(L}_{x},{V}/{L_{x}})\). Any other
geometric parameter of the system will also be characterized by a unique
distribution function. This includes the aspect ratio
\(\tau = L_{x}/L_{y}\), for which the distribution function
\(P_{V}(\tau)\) represents the relative frequency of seeing
conformations with different \(\tau\). This function is directly
accessible in simulations.

Our goal is to evaluate \(P_{V}(\tau)\) and then try to reproduce it in
simulations at a constant volume. Using \({P(L}_{x},L_{y},\sin\alpha)\)
as the starting point, \(P_{V}(\tau)\) can be evaluated as
$P_{V}(\tau) =  \left\langle \delta\left( \tau - {L_{x}}/{L_{y}} \right) \right\rangle_{V}$,
where the brackets \(\langle \cdots \rangle_{V}\) indicate a statistical average over
sub-ensemble with fixed $V$. The constant-volume constraint can be
imposed through another delta function, leading to the following
expression in terms of the unbiased distribution function:
\begin{equation}
P_{V}(\tau)\sim\ \int_{}^{}{\rd L_{x}\rd L_{y}\delta
\left( V - L_{x}L_{y}\sin\alpha \right)
\delta(\tau - L_{x}/L_{y})}\re^{- \beta {\mathcal P}L_{x}L_{y}\sin\alpha}\ 
Q(N,T,L_{x}L_{y}\sin\alpha;L_{x}/L_{y}).
\end{equation}
This  integral can be evaluated in two steps. First, a change of variables $y=L_x L_y \sin\alpha$   
helps to eliminate the integration over  $L_y$ while keeping the other variable constant. The result is: 
\begin{equation}
P_{V}(\tau)\sim\ \int_{}^{}{\rd L_{x}
\delta\left(\tau - \frac{L_{x}^{2}\sin\alpha}{V}\right)}
\frac{1}{L_x \sin\alpha}
\re^{-\beta {\mathcal P} V}
Q(N,T,V; {L_x^2 \sin\alpha}/{V}).
\label{e50}
\end{equation}
The second integration can be carried out by making a change of
variables $y = L_{x}^{2}\sin{\alpha/V}$, which after
dropping the terms that are independent of \(\tau\) yields:
\begin{equation}
P_{V}(\tau) = \frac{1}{\tau}\ Q(N,T,V;\tau).
\label{e5}
\end{equation}
Let us now require that \(P_{V}(\tau)=P(\tau)\), i.e., the distribution
functions in the NPT and NVT ensembles are equal. It is
easy to see then from equation~(\ref{e1}) that
$\int {Q(N,T,V;\tau) f (\tau)\rd\tau }
= \int {P(\tau)\rd\tau }
= \int {P_{V}(\tau)\rd \tau } 
= \int {Q(N,T,V;\tau)({1}/{\tau}) \rd \tau}$,
indicating that \(f(\tau) = 1/\tau\). In other words, we arrive at the
conclusion that new aspect ratios in constant-volume simulations should
be drawn from the \(1/\tau\) distribution in order to reproduce the
result of the NPT simulations. Furthermore, it is seen from equation~(\ref{e5})
that free energy
\(\beta F(\tau) = -\log{\left[ Q(N,T,V;\tau) \right]} \) associated
with the degree of freedom \(\tau\) can be computed as
\(\beta F(\tau) = -\log{\left[ \tau P_{V}(\tau) \right]}\).
Therefore, the aspect ratio \(\tau\) is not suitable for the computation
of free energy differences directly from the distribution function. In
other words, 
\( \ds \beta \Delta F = \beta\left[ F\left( \tau_{1} \right) - F\left( \tau_{2} \right) \right] 
\neq \log{\left[ { P_{V}\left( \tau_{2} \right)}/{P_{V}\left( \tau_{1} \right)} \right]}\),
where \(\tau_{1}\) and \(\tau_{2}\) are some values defining two
macroscopic states. It is easy to show, however, that
\(P_{V}(z) = \tau(z)P_{V}(\tau(z))\) is the distribution function for a
new variable \(z = \log(\tau)\). Thus, in terms of this variable
\(\beta F\left[ \tau(z) \right] = {\beta F(z) = -\log}{\left[ \ \tau(z) P_{V}(\tau(z))\right] = 
-\log[P_{V}(z)]}\),
making $z$ the proper order parameter associated with the geometry
parameter \(\tau\). In the Appendix we show that \(1/\tau\) sampling law
also applies in the three-dimensional space.

\subsection{ Simulation algorithm}

Given that the sampling law is known, how does one conduct a
constant-volume simulation with the variable shape? Let us first point out
the following auxiliary results. The lengths of the box can be expressed
in terms of \(\tau\) and volume when the cell angle is considered
constant: \(L_{x} = \sqrt{{V \tau}/{\sin\alpha}}\) and
\(L_{y} = \sqrt{{V}/{(\tau \sin\alpha)}}\). Given that the volume is
fixed, one can find the appropriate distributions for the lengths using
the standard identity: \(P_{eq}(x)\rd x = P_{eq}(y)\rd y\),
where \(y(x)\) is some function of \(x\) and \(P_{eq}(x)\) is the
distribution function of this variable. It can be shown that they are
given by the same expression as for \(\tau\):
\(f\left( L_{\nu} \right)\sim 1/L_{\nu}\), where \(\nu = x,\ y.\) In other
words, the sampling distributions of the size in $x$ and~$y$
directions obey the same law. This result is of fundamental importance.
The invariance with respect to the swap of $x$ and $y$
coordinates is central to simulations of condensed matter (with the
exception of cases where external fields are imposed that break the
symmetry). Any algorithm that violates this condition should be
considered flawed. For instance, uniform sampling of \(\tau\) 
assumes that
\(f(\tau)\sim \mathrm{const}\), and so one can find that
\(f\left( L_{x} \right)\sim L_{x}\) while
\(f\left( L_{y} \right)\sim{1/L}_{y}^{3}\). Both distributions are
incorrect and will lead to a bias in the sampled ensemble. Another point
that should be made regarding the \(1/\tau\) law concerns its
interpretation. Since small \(\tau\)'s correspond to narrow boxes while
large ones --- to wide boxes, the decline of the distribution function
with \(\tau\) may seem to indicate that the balance between the two
types of boxes is broken. This impression, however, is misleading and
the number of generated narrow and wide boxes is actually the same. The
number of the former can be estimated as
\(f(\tau){\Delta L}/{L_{y}}\), where \(\Delta L\)
is some small range in which \(L_{x}\) is allowed to vary. If, instead,
one varies \(L_{y}\), the same number of boxes should be
\(f(\tau)({L_{x}\mathrm{\Delta}L}/{L^{2}_{y}})\). Now, let us swap
$x$ and $y$ coordinates in the last expression and obtain
\(f\left({1}/{\tau} \right)({L_{y} \Delta L}/{L^{2}_{x}})\).
This operation is expected not to affect the number of boxes, so one should
find that
\(f(\tau){\Delta L}/{L_{y}} = f\left( {1}/{\tau} \right)({L_{y}\Delta L}/{L^{2}_{x}})\).
After some rearrangement, it follows that the condition for the balance
between narrow and wide boxes is
\(f(\tau)\tau = f\left({1}/{\tau} \right){1}/{\tau}\). It is
easy to see that the derived \(1/\tau\) formula satisfies this condition,
proving that the symmetry between different shapes is preserved.

Up to this point we implicitly assumed that \(\tau\) can be changed
by changing either \(L_{x}\) or \(L_{y}\). It is easy to see, however,
that when the volume of the box \(V = L_{x}L_{y}\sin\alpha\) is kept
fixed \({\tau = L}_{x}^{2}\sin{\alpha/V}\). Thus, it is possible to
change the aspect ratio also by changing the cell angle \(\alpha\). It
is clear that the distribution law should not depend on how \(\tau\) is
changed. We conclude, therefore, that the \(1/\tau\) function should also
apply for the angle moves. It is also clear from the expression of the
volume that the term \(\sin\alpha\) can be treated on the same footing
as that of \(L_{y}\). Thus, one can bypass the derivation and immediately conclude
that the sampling probability
\(P(\sin\alpha)\sim{1}/{\sin\alpha}\). The
pertinent order parameter for the cell angle is
\(z = \log(\sin\alpha)\). Accordingly, free
energy associated with \(\alpha\) can be computed as
\( \ds \beta F(\alpha) = - \log\left[ P_{V}(\alpha) ({\sin\alpha}/{\cos\alpha}) \right]\).

It is convenient to combine the two types of moves, one in which
\(L_{x}\) and \(L_{y}\) change simultaneously and one in which
\(\sin\alpha\) changes together with either \(L_{x}\) or \(L_{y}\), in
one algorithm that consists of two steps:
\begin{enumerate}
  \item 
A new value for \(L_{x}'\) is generated from \({1/L}_{x}\)
distribution.
  \item 
A decision is made randomly with equal probability about which step
is to take next: a) new \(L_{y}' = {V}/({L_{x}'\sin\alpha})\),
or b) new \(\sin\alpha' = {V}/({L_{x}'L_{y}})\). The
coordinates of the particles are appropriately scaled by
\(L_{x}'/L_{x}\) and \(L_{y}'/L_{y}\) in the $x$ and $y$
directions in trial moves. Changes of the cell angle \(\alpha\) do not
affect the coordinates.
\end{enumerate}
Generating \(1/\tau\) distributions can be achieved in a variety of
ways. Probably, the easiest is the Metropolis importance
sampling~\cite{r1}. It consists in using a uniformly
distributed random variable to generate a trial~\(\tau'\) that is
then accepted with probability \(1/\tau'\).

\section{Model and methods}
\label{sec:Methods}
\subsection {Mathematical model}

\begin{figure}[htbp] 
\centering 
\includegraphics[scale=0.75]{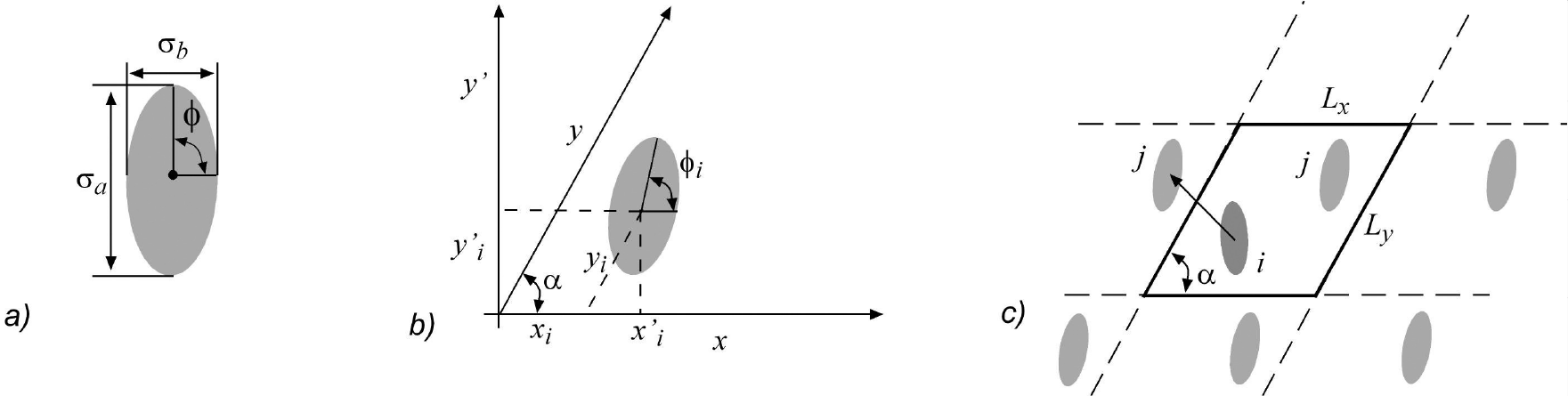} 
\caption{ 
Details of the system studied in this work. (a) Hard-core ellipses
with aspect ratio
$\kappa = {\sigma_a}/{\sigma_b}$.
(b) Skew reference frame used in simulations. 
Each particle is characterized by two translational and one 
rotational variable. (c) Simulation box under periodic boundary
conditions.
} 
\label{f4} 
\end{figure}

We test the designed algorithm for the system of
hard-core ellipses. 
Ellipses have long and short half-axes
\(\frac{1}{2}\sigma_{a}\) and \(\frac{1}{2}\sigma_{b}\), respectively. 
Aspect ratio 
is defined as 
\(\kappa = {\sigma_{a}}/{\sigma_{b}} > 1\).
It is 
a key parameter
determining the general behavior of the system.
The geometrical
details are explained in figure~\ref{f4}~(a).

There are $N$ particles in the system. For the sake of generality we use a skew reference
frame.
Each particle is characterized by three degrees of
freedom: two coordinates \(x_{i},\ y_{i}\) and one rotation angle
\(\phi_{i},\ i = 1,N\), see figure~\ref{f4}~(b) for illustration.
In
addition to the skew coordinates, one can also assign Cartesian
coordinates \(x',\ y'\) to each particle; the two are related as
follows:
\begin{equation}
\left\{
\begin{array}{c}
x' = x + y\cos\alpha, \\
y' = y\sin\alpha, \\ 
\end{array}
\right. \qquad
\left\{
\begin{array}{c}
x = x' -y' \frac{\cos\alpha}{\sin\alpha}, \\
y = y' \frac{1}{\sin\alpha}. \\
\end{array}
\right.
\label{e6}
\end{equation}
The volume elements of the two coordinates systems are related through
the Jacobian of transformation
\(\rd V = \rd x'\rd y' = \frac{\partial (x',y')}{\partial (x,y)} \rd x \rd y= \sin\alpha \rd x \rd y\). 
The particles are placed in a box with sides
\(L_{x}\) and \(L_{y}\) and an angle \(\alpha\) between them under
periodic boundary conditions, as shown in figure~\ref{f4}~(b)--(c).
The skew simulation boxes are designed to
accommodate lattices of all possible types.
The minimum image convention is applied to
compute the interaction energy. As illustrated in figure~\ref{f4}~(c), certain
particle $i$ interacts with the closest periodic image of particle
$j$. The distance between the two particles is computed as
$r_{ij}=\sqrt{x_{ij}^2+y_{ij}^2 + 2 x_{ij} y_{ij}\cos\alpha}$,
where \(x_{ij} = x_{i} - x_{j}\) and
\(y_{ij} = y_{i} - y_{j}\). The usual rule for computing the
shortest distance is used: if \(x_{ij} > L_{x}/2\), then it is
replaced by \(x_{ij} - L_{x}\). Similarly, if
\(x_{ij} < - L_{x}/2\), then it is replaced by
\(x_{ij} + L_{x}\). The same transformations are applied to the
coordinate $y$. The Cartesian coordinates of the vector
\(\rd\vec{r}'\) connecting two particles are
\(\rd x' = x_{ij} + y_{ij}\cos\alpha\) and
\(\rd y' = y_{ij}\sin\alpha\). Together with the rotation angles
\(\phi_{i}\) and \(\phi_{j}\), they are used to determine whether
two ellipses overlap~\cite{r6}.
The density of the system is
reported in reduced units \(\rho = ({N}/{V})\ ({1}/{\rho_{\max}})\)
where \(\rho_{\max} = ({2}/{\sqrt{3}}) [{1}/({\sigma_{a}\sigma_{b}})]\)
is the density of the maximally compact lattice configuration.

\subsection{ Relative free energy }

To measure free energy difference between prospective lattice
structures, we use the properly defined order parameters. As discussed in the section~\ref{sec:Theory},
one such parameter is $\log(\tau)$, where $\tau=L_x/L_y$. The ratio of the side lengths is measured directly
in simulations and then binned to compute the associated distribution.

\begin{figure}[htbp] 
\centering 
\includegraphics[width=1.0\textwidth]{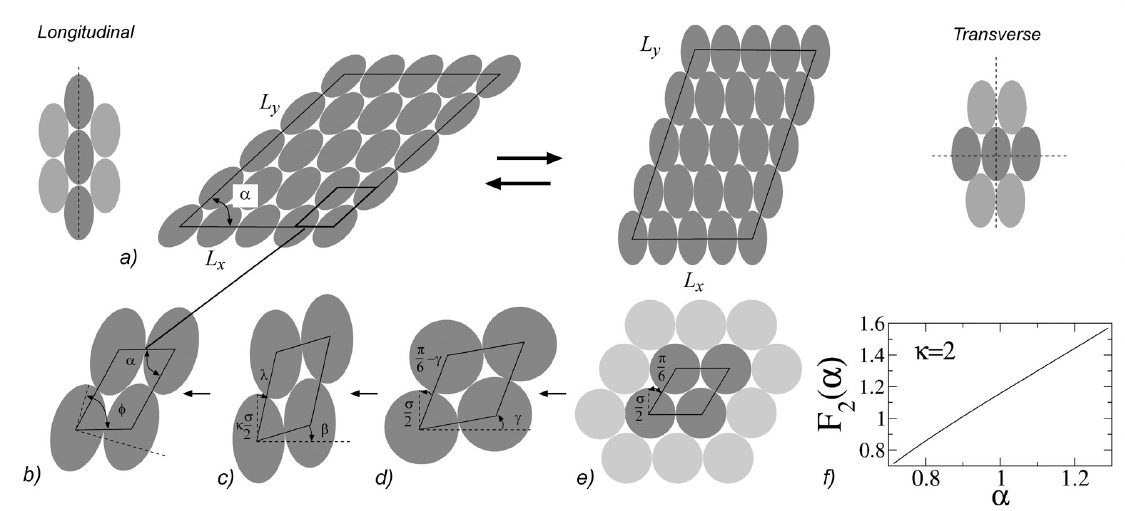} 
\caption{ 
Geometrical details of the simulation cell employed in this study.
(a) Geometry of the cell in the longitudinal and transverse lattice states.
(b)--(e) 
A step-wise procedure to establish one-to-one correspondence between the
cell angle $\alpha$ and the rotation angle of ellipses $\phi$ in the 
close-packed configurations. 
At the first step, e), a minimal non-rectangular box is selected to represent 
the infinite lattice of hard disks with diameter $\sigma$. At the second step, d), the 
box is rotated counterclockwise by an angle $\gamma$. At the third step, c), the 
vertical axis is stretched by the magnitude $\kappa$. At 
the last step, b), the box is rotated clockwise by the angle $\beta$ necessary 
to align the lower side with the horizontal axis. Panel e) --- dependence of $\phi$ on 
$\alpha$ for $\kappa$=2.
} 
\label{f5} 
\end{figure}

Another order parameter that will be utilized is $\alpha$. The rational 
why the cell angle
can be employed as an order parameter is as follows.
The system of ellipses can exist either 
as a fluid or a solid lattice phase, depending on the density~\cite{r3}.
At \(\kappa \lesssim 1.5\)~\cite{r4,r5}, ellipses in the solid phase are 
not aligned with one another,
making the so-called plastic lattice~\cite{r6}. 
At larger
aspect ratios, the particles are aligned similarly to nematic fluids.
An example of the simulation cell 
in which ellipses are aligned 
is shown in
figure~\ref{f5}~(a). The depicted close-packed conformation can 
be generated from
the minimal motif that contains 4 ellipses. It is 
shown in panel~(b) and also highlighted in the right-hand lower
corner of the cell. 
This motif  can be obtained from a close-packed configuration of disks by 
a sequence of unique steps.
Let us assume that the 
diameter of the disks is $\sigma$ and 
the angle that the left side of the initial box makes with the vertical axis is
$\piup/6$, see 
figure~\ref{f5}~(e).
Let us rotate the box counterclockwise by an angle $\gamma$, as shown in
figure~\ref{f5}~(d).
After that let us stretch the box by the amount $\kappa>1$ in the
vertical direction. As a result, disks are transformed into ellipses. The
length of the long axis of the ellipses becomes $\kappa \sigma$, as shown in figure~\ref{f5}~(c). Simultaneously, stretching changes the angle that the lower side of
the box makes with the horizontal axis, as figure~\ref{f5}~(c) illustrates.
While initially it was $\gamma$, now the angle becomes
\begin{equation}
\beta = \cos^{- 1} \left(\frac{\cos\gamma}{\sqrt{\cos^{2}\gamma + \kappa^{2}{\sin^{2}\gamma}}}\right).
\label{e7}
\end{equation}
The angle between the left-hand side and the vertical axis also changes.
Immediately following the rotation, it is $\piup/6 - \gamma$ but after stretching
it becomes
\begin{equation}
\lambda = \cos^{- 1}\left[\frac{\kappa \cos({\piup}/{6} - \gamma)}{\sqrt{\sin^{2}\left({\piup}/{6} - \gamma \right) + \kappa^2{\cos^{2}\left({\piup}/{6} - \gamma \right)}}}\right].
\label{e8}
\end{equation}
As can be seen from figure~\ref{f5}, physically distinct configurations are
generated when $\gamma$ changes between 0 and $\piup/6$. All other values lead to
redundant configurations. To characterize the state of each ellipse in a
specific lattice state, one can use, for instance, the angle that the long
axis makes with the horizontal axis, $\phi$. As can be seen from figure~\ref{f5}~(b)
\(\phi = {\piup}/{2} - \beta\). The geometry of the box, on the
other hand, can be uniquely specified by its angle $\alpha$ (given that the
aspect ratio 
$\kappa$ is fixed). Both angles, $\phi$ and 
$\alpha$, are functions of $\gamma$, so
there is only one independent variable that fully defines the
close-packed structure. All other parameters can be
expressed as functions of the chosen variable via relations~(\ref{e7}) and~(\ref{e8}). For instance, the
rotation angle $\phi$ can be cast as a function of $\alpha$:
\begin{equation}
\phi = F(\alpha;\kappa),
\label{e9}
\end{equation}
where $\kappa$ is the aspect ratio. For the purpose of illustration,  figure~\ref{f5}~(e) shows 
this function for $\kappa=2$, \(F_{2}(\alpha)\).

It can be shown that $\alpha$ varies 
in the range
\( \{\alpha_{\rm {min}}=\cos^{- 1}[\sqrt{{3}/({3 + \kappa^{2}})}],\
\alpha_{\rm {max}}={\piup}/{2} - \cos^{- 1}[\sqrt{{3\kappa^{2}\ }/({3 + \kappa^{2}})}] \} \),
while the corresponding limits of the rotation
angle $\phi$ are
$\{\phi_{\rm {min}}={\piup}/{2} - \cos^{- 1}[\sqrt{{3\ }/({3 + \kappa^{2}})}],\ \phi_{\rm {max}}= {\piup}/{2} \}$.
As $\alpha$ changes, the structure of the lattice changes with it, 
creating 
the manifold of an infinite number of lattices  differing from one another by the rotation angle $\phi$.
This is in contrast to the system of disks, which makes  a single
hexagonal lattice at high densities.
As will become clear below, the lattice types 
corresponding to the extreme points have a special meaning. At $\alpha=\alpha_{\rm {min}}$,
see figure~\ref{f5}~(a), the lattice can be viewed as being assembled from vertical lines of ellipses
that make contact with one another through the pole at the short side of the particle.
The ellipses are aligned along the line connecting their centers. Consequently, the
corresponding lattice type is termed longitudinal~(L). 
The lattice observed at $\alpha=\alpha_{\rm {max}}$ can be assembled from horizontal lines of ellipses which
contact each other through the pole at their long side. The orientation of particles is
perpendicular to the line connecting their centers, so  
the corresponding lattice type is 
called transverse~(T).
In simulations,
the longitudinal lattice can be transformed into the transverse lattice and vice versa by varying the
angle $\alpha$. If~$\alpha$ is treated as a variable, the transition will happen spontaneously. The
system in this case will sample various types of lattices in the course of a single simulation, 
thereby enabling the 
computation 
of their statistical weight.
A major difficulty associated with this approach is convergence. To collect sufficient
statistics for the distribution function $P(\alpha)$, the system should visit
different lattice types a large number of times, which may be difficult if the free energy difference 
between the concerned lattices is large.
To improve the convergence, in this study we employ the method of umbrella sampling~\cite{r7}.
The range $[\alpha_{\rm {min}},\alpha_{\rm {max}}]$ is broken equidistantly into a number of bins, or windows,
each assigned a distinct value of $\alpha$. Harmonic potential is applied 
to bias sampling in simulations to 
the vicinity of each window.
The unbiased distribution $P(\alpha)$ is reconstructed by collecting information on $\alpha$
sampled in all windows.
The order parameter relevant for the cell angle is $\log(\sin\alpha)$.

\subsection{Method of Einstein crystal }

We also compute free energy differences by the method of Einstein crystal~(EC)~\cite{r11,r12}.
The key idea of this method is to obtain free energy
of the lattice states of interest relative to a common reference state. 
The free energy of the reference state then drops when the free energy difference is taken.
The reference
state is modelled by the harmonic Hamiltonian $U_{\rm H}$. Free energy with respect to 
this model
is computed by thermodynamic integration with the help of a coupling constant
$\lambda$. How this scheme works in skew coordinates is described in detail in the Appendix. The main
formula
we use to compute the free energy difference between lattice states characterized by different $\alpha$'s
is:
\begin{equation}
\Delta F(\alpha)=
- k T (N-1) \log (\sin\alpha)
+ {\displaystyle \int_0^1}\langle\Delta U\rangle_{\lambda} \rd \lambda.
\label{ecm}
\end{equation}
Here,
$\langle\Delta U\rangle_{\lambda}$ is the average of the potential energy difference $\Delta U = U_{\rm {HS}} - U_{\rm H}$
computed in simulations driven by the ``hybrid'' Hamiltonian $U(\lambda)=U_{\rm H} + \lambda \Delta U$. 
The actual potential energy of the hard-ellipse system is $U_{\rm {HS}}$ and the simulations are
performed 
at a fixed volume $V$ and 
cell angle
$\alpha$
in the reference frame associated with the center of mass. 
As $\lambda$ is varied between 0 and 1, the Hamiltonian $U(\lambda)$ is transformed from
$U_{\rm H}$ to $U_{\rm {HS}}$.
As the harmonic potential gradually becomes weaker, 
the integral in equation~(\ref{ecm}) 
reports on the spatial extent by which the system is allowed to deviate from the
initial configuration, thus providing a measure of the configurational freedom.

The reference Hamiltonian
\[U_{\rm H}=\displaystyle \sum_{i=1}^{N}\left\{ \frac{\gamma_T}{2}\left[ (x_i-x_i^0)^2+(y_i-y_i^0)^2 \right]
+ \frac{\gamma_R}{2}(\phi_i-\phi_i^0)^2 \right\}
\]
contains a set of coordinates
$x_i^0, y_i^0,~i=1,N$ with respect to which the
free energy is evaluated. 
As a reference we chose a lattice configuration $x_i^0=(j-0.5) L^E_x/n, j=1,n$ and 
$y_i^0=(k-0.5) L^E_y/n,\ k=1,n$; 
${ i= \sum_{j'=1}^{j} \sum_{k'=1}^k }1$, $n \times n=N$.
Here, $ L^E_x$ and $ L^E_y$ are the dimensions of the simulation cell in $x$ and $y$ direction,
$\phi_i^0$ are initial angles and 
$\gamma_T$ and $\gamma_R$ are adjustable spring constants. 
It is easy to show that ${N}/{\rho}= \tau {L^E_y}^2 \sin\alpha$, where $\tau={L^E_x}/{L^E_y}$
is the ratio of the $x$ and $y$ dimensions. Thus, $L^E_y$ is uniquely defined by $N$, $\rho$, $\alpha$
and $\tau$. 
Parameter $\tau$ was extracted 
from
constant-volume simulations 
as the average over all sampled cell side ratios.
The integral in equation~(\ref{ecm}) was evaluated by numerical quadrature. 
In order to reduce the variation seen in $\langle\Delta U\rangle_{\lambda}$,
a non-uniform transformation $\zeta(\lambda)$ was applied before integration.
A total of 
33 grid points, $\zeta_i, i=1,33$, were generated non-uniformly between $\zeta(0)$ and $\zeta(1)$. 
The points were chosen so as to
maintain a constant level of the numerical 
integration error, which, after optimization, we judge to be negligible. 
The statistical error was estimated by performing 5 independent measurements for each state point
from which standard deviation was extracted.

\subsection{Numerical details}

The performed MCVS simulations consist of two types of moves. The first type is the
regular MC steps in which particles are randomly displaced and rotated.
The maximum magnitude of these steps are adjusted to achieve greater
than 30\% acceptance. The second type of moves are changes of geometry.
They are attempted randomly with 10\% overall probability. Box lengths
and box angle are updated as described in the previous section.
Parameters of these moves are adjusted to achieve greater than 50\%
acceptance rate. Changes of box lengths are accompanied with the
appropriate scaling of particle coordinates. The algorithm of
Vieillard-Baron~\cite{r6} is used to determine if two ellipses
overlap.

To perform the umbrella sampling simulations,
the range {[}\(\alpha_{\rm {min}}\),\(\ \alpha_{\rm {max}}\){]} was divided
equidistantly into 50
windows,\(\ \alpha_{i} = \alpha_{\rm {min}} + \Delta (i - 1),\ \ \Delta = ({\alpha_{\rm {max}} - \alpha_{\rm {min}}})/{49},\ i = 1,50\).
Simulations were performed in each window with an additional harmonic
potential
\(U_{h} = \frac{1}{2}\epsilon\left( \alpha - \alpha_{i} \right)^{2}\)
applied. The strength of the potential~\(\epsilon\) was set such that
the adequate overlap of the histograms at the neighboring windows was
achieved. The unbiased distribution \(P_{V}(\alpha)\) was obtained at
the end of the simulations by the multiple-histogram reweighting method~\cite{r9,r10}.

\section{Results}

We test the proposed algorithm in simulations of the system of hard-core ellipses.
For the sake of comparison, 
we consider the scheme in 
which $\tau$ is sampled uniformly, i.e., the  $\tau$-sampling algorithm,
in addition to the proper 1$/\tau$ sampling law.
It is found that proper sampling is essential
when the parameter controlling the geometry of the cell varies in a wide
range. In other cases, different sampling methods lead to
indistinguishable results.

\subsection{Solid plastic phase }

\begin{figure}[htbp] 
\centering 
\includegraphics[width=1.0\textwidth]{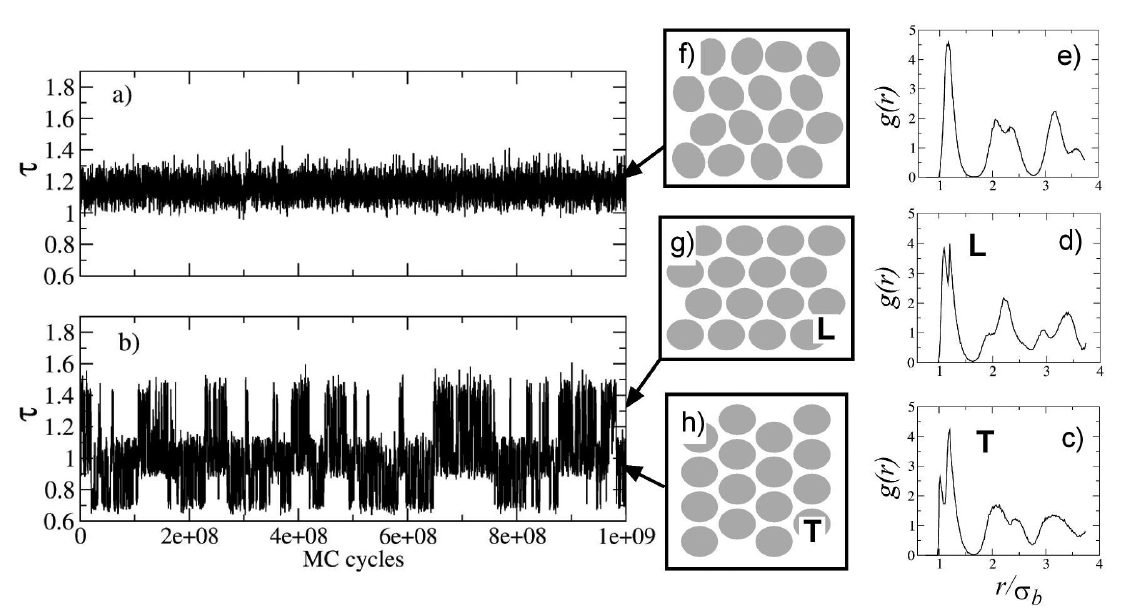} 
\caption{ 
 Time traces of the aspect ratio $\tau$ obtained for
models with $\kappa=1.2$ in which a) ellipses have
different rotation angles and b) --- the same rotation angle. Panels f),
g) and h) show cartoon representation of lattices in which ellipses are
not aligned, or are aligned in the longitudinal (L) and transverse (T)
structures, respectively. The pair distribution functions corresponding
to these lattices are shown in panels e), d) and c).
} 
\label{f6} 
\end{figure}

The first test was performed for a system with \(\kappa = 1.2\). We used
rectangular boxes and set the density at \(\rho = 0.83\) which is high
enough to trigger the transition into the crystalline state yet low enough
to preclude aligning of the ellipses. Under these conditions, the system
makes the so-called plastic lattice in which particles occupy lattice
sites but are capable of rotating by all 180 degrees. A time trace of \(\tau\)
observed in simulations of this system is shown in figure~\ref{f6}~(a). The
aspect ratio varies 
in the
range $[1.0,1.3]$ 
around an average of~\textasciitilde1.15. 
At all times, the system occupies a lattice
configuration with the distance between neighboring sites
\(1.17 \sigma_{b}\) while the ellipses are capable of adopting any angle
\(0 < \phi < \piup\). A cartoon representation of this structure and its
pair distribution function computed for the centers of the ellipses 
are shown in figure~\ref{f6}~(f) and figure~\ref{f6}~(e),
respectively. The lack of alignment between particles is immediately
apparent.

We find that the alignment (or rotational degrees of freedom more
generally) plays a critical role in determining the structure of the
lattice. When it is introduced manually by constraint 
\(\phi_{i} = \phi,\ i = 1,N\) where \(\phi\) is the common angle of all particles which is allowed
to change, the
lattice splits into two sub-lattices with a distinct structure. One
sub-lattice displays contacts between
neighboring ellipses going through the pole at the long side and
can be recognized as the transverse lattice.  A cartoon representation of this lattice 
is shown in figure~\ref{f6}~(h). 
The second sub-lattice is the 
longitudinal lattice, 
exhibiting closest contacts between the neighboring
ellipses going through their short side, as illustrated in figure~\ref{f6}~(g).
All other lattice configurations correspond to free energy maxima and are not
directly observable. 
It is not clear if this is a genuine property of the system or  
an artifact resulting from the use of rectangular box geometry,
suppressing all structures other than the mentioned two.
For the demonstration purposes, we
considered a small system
consisting of $N=16$ particles making $4 \times 4$ lattice, so as to enable
spontaneous T-L transitions. The time trace of
\(\tau\) obtained in simulations of this system is shown in figure~\ref{f6}~(b). It is seen
that \(\tau\) changes in a discontinuous manner among four different
values, indicating that the geometry of the box is specific to the
lattice type that it contains and suggesting that the aspect ratio can be
used as a structural parameter to distinguish between T and L states.
Pair  distribution functions \(g(r)\) obtained for each lattice type, shown in 
figures~\ref{f6}~(c),~(d),
demonstrate that
the two 
lattices have a noticeably different structure.
The most obvious
differences concern the first coordination shell. Compared to the model
with full particle rotations, \(g(r)\) in the T and L states is split
into two sub-maxima. The position of the first sub-maximum in T
conformations is at \(1.01 \sigma_{b}\) while in L conformations it is
at \(1.1 \sigma_{b}\), demonstrating that the particles in the transverse
arrangement are capable of approaching each other at shorter distances. This
is understandable, given how ellipses are stacked in the two structures,
see figure~\ref{f6}~(g) and~(h). The second sub-maximum appears at
\(1.2 \sigma_{b}\) in both states. The second and third coordinate
shells also show significant differences between T and L conformations.

\begin{figure}[htbp] 
\centering 
\includegraphics[width=0.6\textwidth]{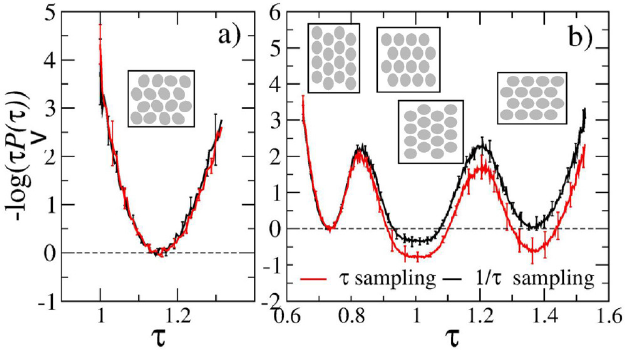} 
\caption{(Colour online) 
 Free energy profile obtained for ellipses with
$\kappa = 1.2$ in simulations where particles have full
rotational freedom, a), and where particle rotation angles are rigidly
coupled, b). Minima seen for the constrained system correspond to
various stable structures consistent with particular~$\tau$
such as transverse and longitudinal states. 
The results of 
two sampling methods are as labeled.
The $\tau$-sampling
method is seen to shift artificially the balance in the statistical weight toward
wider boxes.
Error bars were estimated from 5 independent trajectories.
} 
\label{f7} 
\end{figure} 

So how well does the proposed algorithm reproduce the statistics of the
different conformational states? Figure~\ref{f7} shows the free energy profile
\(\beta F(\tau) = - \log[\tau P_{V}(\tau)]\) obtained for the system
where all particles are free to rotate --- panel a), and for the
rotationally-constrained system --- panel b). As can be expected from
figure~\ref{f6}, a single minimum is seen for the unconstrained system and four
minima are present for the constrained one. The specific aspect ratios of the minima
are \(\tau=0.7\), $0.95$, $1.05$ and~$1.4$. The two extreme values, $0.7$ and
$1.4\approx1/0.7$, correspond to the longitudinal state. The two minima in the
middle are characteristic of the transverse state; they appear to merge 
in the graph because of being closely spaced. 
Conformations with \(\tau=0.95\) and $1.05$, and those with \(\tau=0.7\)
and $1.4$, are related by a 90-degree rotation of the reference frame
(note that no such transformation ever takes place in the simulations).
Corresponding to the same state, they should exhibit the same free
energy. It is a crucial test for the algorithm to reproduce this
behavior. It is seen from figure~\ref{f7}~(b) that
\(\beta F(0.95) = \beta F(1.05)\) and \(\beta F(0.7) = \beta F(1.4)\)
within the error bars, so the proposed \(1/\tau\) algorithm successfully
passes the test. By contrast, the \(\tau\)-sampling algorithm predicts
that \(\beta F(0.95) = \beta F(1.05)\) but
\(\beta F(1.4) < \beta F(0.7)\) and the difference is statistically
significant. The greater population of \(\tau = 1.4\) state can be
explained by the greater statistical weight assigned to conformations
with larger \(\tau\) by an algorithm in which the sampling probability is
flat compared to the case when it declines with $\tau$. Furthermore,
even though the \(\tau\)-sampling method generates the same free energy
for \(\tau=0.95\) and $1.05$, the value it predicts is twice as large as
that of the proper algorithm. Again, the difference is statistically
meaningful, which leads us to the conclusion that this method is not
capable of reproducing accurately the inter-state statistics for the
constrained model under the considered conditions.

By contrast, figure~\ref{f7}~(a) shows that both sampling algorithms lead to the
same results, within error bars, for the unconstrained system. It
follows, therefore, that the performance of geometry sampling algorithms 
strongly depends on the system
under study, more specifically on the range in which the geometry
parameter is varied. If the range is narrow, as in figure~\ref{f7}~(a), the use
of the proper algorithm does not make much difference. The average
\(\tau\) and the shape of the distribution function are generally well
reproduced. The algorithm matters, however, when \(\tau\) changes in a wide range,
as shown in figure~\ref{f7}~(b). Improper sampling leads to significant
distortions in \(P_{V}(\tau)\) that may ultimately affect the free
energy estimates.

\subsection{Free energy difference between distinct lattice types }

When \(\kappa\) is increased beyond \textasciitilde1.5, the ellipses in
the crystalline state become aligned~\cite{r4,r5}. What is the statistical weight of 
lattice types with different ellipse orientations?  To find that out, we considered
the system with $\kappa=4$.
The cell angle $\alpha$ for this system varies in the range [0.41, 1.43], which is much wider
than the range 
[0.96, 1.12] appropriate for
\(\kappa = 1.2\). This improves the chances of observing a quantitative difference between
the results of using different sampling methods.
We considered the cells with 6 rows and 6 columns of ellipses
because smaller systems failed to produce stable lattices. Since spontaneous transitions among different
lattice types were not observed for the considered system, we had to employ
umbrella sampling method to compute the distribution function of the structural order parameter $\alpha$.
The details are provided in the section~\ref{sec:Methods}. 
The density of the system was set at \(\rho = 0.95\) which is much
higher than the density of the fluid-solid transition. 

\begin{figure}[htbp] 
\centering 
\includegraphics[width=0.5\textwidth]{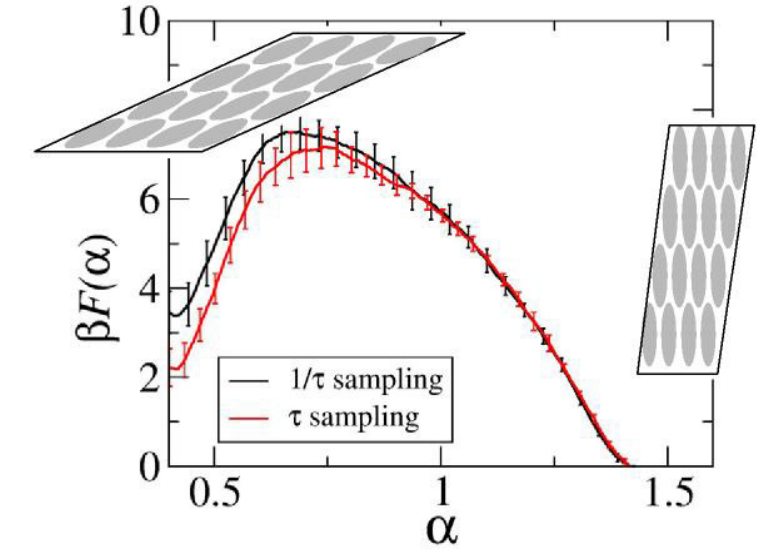} 
\caption{(Colour online) 
 Free energy profile obtained for ellipses with
$\kappa = 4$. Cartoons show longitudinal and transverse states
corresponding to the two minima. Results of the two sampling
methods are as labeled.
} 
\label{f8} 
\end{figure}

Free energy profile
$\beta F(\alpha) = - \log \left[ P_{V}(\alpha)\frac{\sin\alpha}{\cos\alpha} \right]$ obtained
in simulations using the proposed sampling algorithm is shown in figure~\ref{f8} by black
line.
It has two minima with the lowest free energy
corresponding to the transverse and longitudinal states. 
This agrees with our results for the constrained model of \(\kappa = 1.2\),
suggesting that the presence of 
two identified stable states is not an artifact of the used
cell geometry.
The free energy of the longitudinal
state is \(\beta \Delta F = 3.5 \pm 0.5\) higher than that of the
transverse state. 
The corresponding figure obtained by the $\tau$-sampling method is 
\(\beta \Delta F = 2.2 \pm 0.4\). 
It is seen that this method
assigns additional statistical weight to the longitudinal state
with smaller $\alpha$.
This is in line with
our observation from the previous section where we saw states with
larger $\tau$ experiencing more frequent sampling. Indeed, since $\sin\alpha={V}/{\tau L_y^2}$, more sampling
for large $\tau$ means more sampling for small $\sin\alpha$ and, by extension, small $\alpha$.

Free energy difference between the L and T states was also evaluated by the EC method. The geometry of the
cell was determined in simulations using $1/\tau$ sampling. For the longitudinal structure we used
$\alpha=\phi^0=\alpha_{\rm {min}}=0.43$ and $\tau=0.558$. For the transverse state, the corresponding numbers were
$\alpha=\alpha_{\rm {max}}=1.43$, $\tau=0.288$ and $\phi^0=\piup/2$. The initial ellipse orientations $\phi^0$ 
were generated
by $F_4(\alpha)$
for both structures. 
Upon combining the results of 5 independent measurements, the free 
energy difference was evaluated as $\beta \Delta F=
3.2\pm0.2$.
This number is in excellent numerical agreement with the prediction made in the $1/\tau$ sampling simulations,
$3.5\pm0.5$, providing an essential validation for this method.
By contrast, the $\tau$-sampling algorithm generates an incorrect free energy, $2.2\pm0.4$, and the error
is statistically significant.
Based on these observations,
we conclude that the use of the proper
sampling algorithm is essential for the studied system. 

The free energy of the transverse state appears to be lower than that of the longitudinal
state. This could be a genuine physical effect stemming from conformational preferences of
different lattice configurations.
Alternatively, the difference could be a by-product of the small size of the simulation cell.
Simulations 
designed to extract free energy difference as a function of $N$ could help to resolve this ambiguity.
Linear segments in 
$\beta \Delta F (N)$ at large $N$ would indicate genuine free energy difference between the two lattice states.
Any other dependence would signal  
finite-size artifacts.
Which of these two scenarios takes place needs to be answered in careful finite-size scaling studies.

\section{Conclusions}

We introduced an algorithm to perform MC simulations of crystalline
systems in boxes with fixed volume but variable shape. Tests performed
for the system of hard-core ellipses showed that the performance of the
algorithm depends on the range in which the geometrical parameter
characterizing the shape varies. If the range is narrow, which is
probably the case for the majority of crystalline simulations, the use
of the algorithm does not make much difference in comparison with other
\emph{ad hoc} sampling schemes. For wide ranges of the parameters, 
however, the use of the
algorithm may be crucial. In the example of structural transition
between transverse and longitudinal lattices, we find that the error in the
free energy difference due to the inadequate sampling method reaches
40\%. How large it may become, and how wide the corresponding variation
range should be, will depend on the system of interest. However, as a
general rule, the error due to the incorrect sampling should not be
simply neglected.

At the same time, we note that the magnitude of the error declines when
the range of the sampled parameter becomes narrower. This will happen,
among other reasons, when the number of particles~$N$ increases. Thus,
proper finite-size analysis, in addition to yielding important
information about the scaling properties of the studied system, will also
help to combat the errors associated with the inadequate methods of sampling
simulation box shapes.

\section*{Acknowledgements}
The author expresses his utmost gratitude to the men and women of 
the Armed Forces of Ukraine, the National Guard and other law-enforcement
agencies who made 
this study possible by their selfless service and, for many, 
ultimate sacrifice.
The author would also like to thank Gerhard Kahl, Susanne Wagner, Roman Melnyk 
and Andriy Stelmakh for 
stimulating discussions during the time when this study was conceived and completed. 
This study was supported  by 
the National Academy of Sciences of Ukraine, 
project KPKVK 6541230.

\section*{Appendix}

\newcounter{appfigure}
\renewcommand{\thefigure}{A.\arabic{appfigure}}
\let\oldcaption\caption
\def\caption{\stepcounter{appfigure}\oldcaption}

%\newcounter{appequation}
\renewcommand{\theequation}{A.\arabic{equation}}
\renewcommand{\thesubsection}{A.\arabic{subsection}}

\subsection{Algorithm for the three-dimensional space}

\begin{figure}[htbp] 
	\centering 
	\includegraphics[width=0.4\textwidth]{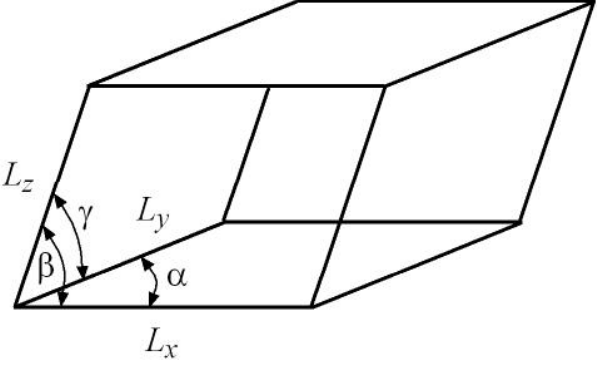} 
	\caption{ 
		Simulation box in 3D is defined by three variables: three box
		lengths $L_x$, $L_y$  and $L_z$,
		and three angles
		$\alpha$, $\beta$ and $\gamma$.
	} 
	\label{f9} 
\end{figure}

It is possible to conduct constant-volume simulations with the changing
shape in the three-dimensional space. Let us assume that the geometry of the
box is defined by six variables, three lengths \(L_{x},L_{y}\) and~\(L_{z}\), and three angles \(\alpha,\beta\) and \(\gamma\), as shown in
figure~\ref{f9}. Let us further assume for now that the box is rectangular,
\(\alpha = \beta = \gamma = \piup/2\). The analogue of the
constant-pressure distribution~(\ref{e3}) in 3D will be a function of the
lengths of the box:
\begin{equation}
P\left( L_{x},L_{y},L_{z} \right)\sim \re^{- \beta {\mathcal P}L_{x}L_{y}L_{z}}\ Q(N,T,L_{x}L_{y}L_{z};\tau).
\label{e10}
\end{equation}

Aspect ratios can be computed for all three sides of the box lying in
\(xy\), \(xz\) and \(zy\) planes. We will try to
change only one \(\tau\) at a time. The corresponding distribution
function, for instance for \(\tau = L_{x}/L_{y}\), can be computed as
follows:
\begin{equation}
P_{V}(\tau)\sim\ \int_{}^{}{\rd L_{x}\rd L_{y}\rd L_{z}\delta\left( V - L_{x}L_{y}L_{z} \right)
\delta(\tau - L_{x}/L_{y})}\re^{- \beta {\mathcal P}L_{x}L_{y}L_{z}}\ Q(N,T,L_{x}L_{y}L_{z};L_{x}/L_{y}).
\label{e11}
\end{equation}
By making the substitution \(y = L_{x}L_{y}L_{z}\) one can eliminate the
integral over \(L_{z}\) obtaining: 
\[
P_{V}(\tau)\sim\ \int_{}^{}{\rd L_{x}\rd L_{y}\frac{1}{L_{x}L_{y}}\ 
\delta(\tau - L_{x}/L_{y})}\re^{- \beta {\mathcal P}V}\ Q(N,T,V;L_{x}/L_{y}).\]
After the second substitution \(y = L_{x}{/L}_{y}\) and elimination of
\(L_{x}\), one arrives at:
\[P_{V}(\tau)\sim\ \re^{- \beta {\mathcal P}V}\int_{}^{}{\rd L_{y}\frac{1}{L_{y}} }\frac{1}{\tau}  Q(N,T,V;\tau) \sim\ \frac{1}{\tau} Q(N,T,V;\tau).\]
This result does not depend on the integration order. Therefore, we find
that the sampling law in three-dimensional space is the same as in the
two-dimensional space.

Since there is no difference with respect to the dimensionality, it is
convenient to use the algorithms developed for 2D and apply them to 3D
boxes. In particular, the idea of sampling one side of the box and then
updating alternatively either another side or the angle between the
sides can be reused. The volume of the box with variable angles is
\[V = L_{x}L_{y}L_{z}\sqrt{1 - {\cos^{2}\alpha} - {\cos^{2}\beta} - {\cos^{2}\gamma} + 2\cos\alpha\cos\beta \cos\gamma}.\]
In MC sampling, one should take alternatively pairs of a side and a
second side or a side and an angle until all such combinations are
exhausted. If the length \(L_{x}\) is updated to \(L'_x\) then
companion steps should be updating \(L_{y}\) or \(\cos\alpha\). The
second possible combination is (\(L_{x}\), \(L_{z})\) or \((L_{x},
\ \cos\beta)\). The third and final combination is (\(L_{y}\),
\(L_{z})\) or \((L_{y}, \ \cos\gamma)\). Specific formulas for how to
update the lengths and angles can be obtained from the condition that
the volume of the cell remains constant. If for instance \(L_{x}\) is
sampled and a trial value is \(L'_x\), then the trial value for the
length in the $y$-direction should be \(L'_y = V/L'_x\) and
the trial angle in the $xy$ plane
\[\cos\alpha' = \sqrt{\cos\beta\cos\gamma + \sqrt{{\cos^{2}\beta}{\cos^{2}\gamma} + 1 - {\cos^{2}\beta} - {\cos^{2}\gamma} - {V^{2}}/{{L'^2_x}L_{y}^{2}L_{z}^{2}}}}.\]
Similar formulas can be obtained for \((L_{x},
\ L_{z},\ \cos\beta)\) and \((L_{y},\ L_{z},\cos\gamma)\)
combinations by permutations.

\subsubsection{Einstein crystal method in skew coordinates}
There are certain caveats associated with the use of EC method  in conjunction with the
skew coordinates, which we explain below. 
To evaluate the Hamiltonian, we need momenta in addition to the coordinates.
These can be obtained using a
general formula
$P_x={\rd K}/{\rd \dot x}$, where
$x$ is the coordinate, $\dot x$ is the
time derivative of the coordinate and $K$ is the kinetic energy.
For Cartesian coordinates, the kinetic energy is known:
$K=\frac{m}{2}( {\dot x}'^2 + {\dot y}'^2)$, where $m$ is the mass of the particle.
It yields the familiar momenta
$P_x'=m {\dot x}'$ and $P_y'=m {\dot y}'$.
Assuming that skew and Cartesian coordinates are related by expression~(\ref{e6})
it follows that
\begin{equation}
K=\frac{m}{2} ({\dot x}^2 + {\dot y}^2 + 2 {\dot x} {\dot y} \cos\alpha),
\label{kinetic}
\end{equation}
thus, one can find the skew momenta:
$P_x={\rd K}/{ \rd \dot x}=m (\dot x + \dot y \cos\alpha) = P_x'$
and 
$P_y= m (\dot y + \dot x \cos\alpha) = P_x' \cos\alpha + P_y' \sin\alpha$.
The
reverse transformation 
$P_x'=P_x$ and $P_y'=\frac{1}{\sin\alpha}(P_y-P_x \cos\alpha)$
allows one to 
compute the Jacobian: 
$\ds { \partial (P_x',P_y')}/{ \partial P_x,P_y}= {1}/{\sin\alpha}$ leading to
the volume element $\rd P_x' \rd P_y'= \frac{1}{\sin\alpha} \rd P_x \rd P_y$. The kinetic energy
$$K=\frac{1}{2 m} \frac{1}{\sin^2\alpha} (P_x^2 + P_y^2 - 2 P_x P_y \cos\alpha)$$
can be found by expressing time derivatives of the coordinates in terms of the momenta
and then substituting the results into equation~(\ref{kinetic}).

Now, let us compute the partition function for one particle:
$$Q(N,V,T)= \frac{1}{1! h^2} \int{ \re^{-\beta H} \rd P_x' \rd P_y' \rd x' \rd y'} = \frac{V}{1! h^2}
\int{ \re^{-\beta K} \rd P_x' \rd P_y' \frac{1}{V} \int{ \re^{-\beta U} \rd x' \rd y'}} = Q_{\rm{id}} \times Q_{\rm{ex}},$$
where we used the property that coordinates and momenta can be
partitioned and \emph{U} is the potential energy of the system. The two
integrals in this expression correspond to ideal $Q_{\rm{id}}$
and excess $Q_{\rm{ex}}$
partition functions giving rise to ideal and excess free energy,
correspondingly. Let us evaluate the ideal part in the skew reference
frame:
\begin{align}
Q_{\rm{id}}&=\frac{V}{1! h^2} \int{ \re^{- \beta K} \frac{1}{\sin\alpha} \rd P_x \rd P_y} \nonumber\\ 
&=
\frac{V}{h^2} \int{ {\rm{exp}} \left\{- \beta 
\frac{1}{2 m} \frac{1}{\sin^2\alpha} \left[P_x^2 + P_y^2 - 2 P_x P_y \cos\alpha\right]
\right\} \frac{1}{\sin\alpha} \rd P_x \rd P_y} =
\frac{V}{h^2} \left( \frac{2 \piup m}{\beta} \right), \nonumber
\end{align}
 which turns out to be exactly the same
as the equivalent expression in the Cartesian coordinates. Extending these calculations to
$N$ particles will produce $Q_{\rm{id}} = ({1}/{N!}) ({V^N}/{h^{2 N}}) ({2 \piup m}/{\beta})^N$,
which for sufficiently large $N$ will give the familiar expression of the free energy
of ideal gas:
$\beta F_{\rm{id}}= N [ \log(\rho) - \log({1}/{\lambda ^2}) -1]$, where 
$\lambda = \sqrt{{\beta h^2}/{ 2 \piup m}}$.
Note that this expression does not depend on $\alpha$, so the ideal part of the
free energy can be neglected when working in the skew reference frame 
if relative free energy is  of interest.
Moving on to the excess partition function, one finds for one particle:
$$Q_{\rm{ex}}= \frac{1}{V} \int{ \re^{-\beta U} \rd x' \rd y' } = \frac{\sin\alpha}{V} 
\int{ \re^{-\beta U} \rd x \rd y }. $$
Generalization to $N$ particles
\begin{equation}
Q_{\rm{ex}}=
\frac{\sin^N\alpha}{V^N}\int{\re^{-\beta U} 
{\displaystyle 
\prod_i} \rd x_i \rd y_i}
\label{exess}
\end{equation} 
gives rise to the excess free energy $\beta F_{\rm{ex}}(N,V,T)=-\log{Q_{\rm{ex}}}$, which is
fully defined by the interaction energy $U(x_1,y_1,...,x_N,y_N)$.

Let us focus on the system of hard disks for now, and introduce a scaling constant 
$\lambda$ that will transform the hard-sphere potential 
$U_{\rm {HS}}$ into the harmonic potential 
$U_{\rm H}={ \sum_{i=1}^{N}} ({\gamma_T}/{2}) \left[ (x_i-x_i^0)^2+(y_i-y_i^0)^2\right]$:
\begin{equation}
U(\lambda)=U_{\rm H} (1-\lambda) + \lambda U_{\rm {HS}}  = U_{\rm H} + \lambda (U_{\rm {HS}}-U_{\rm H}) = U_{\rm H} + \lambda \Delta U, \qquad
\Delta U=U_{\rm {HS}} - U_{\rm H}.
\end{equation}
Here, $x_i^0, y_i^0,~i=1,N$ are the positions of the ideal lattice with respect to which the
free energy is evaluated. 
At $\lambda=0$, 
the system is described by the harmonic potential. 
It constitutes a set of independent harmonic oscillators with 
frequency controlled by the spring constant 
$\gamma_T$.
At $\lambda=1$, 
we obtain the system of our interest --- hard spheres.
The ``hybrid'' potential $U(\lambda)$
defines some fictitious system which physically makes sense only 
at the extreme points of $\lambda$.
The free energy of that fictitious system 
$\beta F_{\rm{ex}}(\lambda)=-\log{Q_{\rm{ex}}}$
can be used to compute the free energy difference between the end points:
$\Delta F = 
F_{\rm{ex}}(\lambda=1) - F_{\rm{ex}}(\lambda=0) = F_{\rm {HS}} - F_{\rm H}$. Since $F_{\rm H}$ is 
 known analytically, this formula allows us to compute $F_{\rm {HS}}$ as a sum 
$F_{\rm {HS}}= F_H + \Delta F$, making the evaluation of $\Delta F$ a key task.
This task can be accomplished by taking the derivative of $F_{\rm{ex}}(\lambda)$ with respect 
to $\lambda$:
\begin{equation}
\frac{\rd \beta F_{\rm{ex}}(\lambda)}{\rd \lambda} = -\frac{1}{Q_{\rm{ex}}(\lambda)} 
\frac{\rd Q_{\rm{ex}}(\lambda)}{\rd \lambda} = -\frac{1}{Q_{\rm{ex}}(\lambda)}
\frac{\sin^N\alpha}{V^N} \int \re^{-\beta U(\lambda)} \beta \Delta U 
\prod \rd x_i \rd y_i = \beta \langle\Delta U\rangle_{\lambda},
\end{equation}
where symbol $\langle...\rangle_{\lambda}$ denotes average in ensemble generated by $U(\lambda)$.
After the integration of the derivative, one obtains:
$$\Delta F={\displaystyle \int_0^1 }
\frac{\rd F_{\rm{ex}}(\lambda)}{\rd \lambda} \rd \lambda = {\displaystyle \int_0^1}\langle\Delta U\rangle_{\lambda} \rd \lambda.
$$
This is the key formula which  
relates the free energy of the system of 
interest $F_{\rm {HS}}$ to the integral obtained in simulations of the hybrid system:
\begin{equation}
F_{\rm {HS}} = F_{\rm H} +  {\displaystyle \int_0^1}\langle\Delta U\rangle_{\lambda} \rd \lambda.
\end{equation}
Provided that $\gamma_T$ is taken sufficiently large, the
partition function in harmonic approximation can be evaluated analytically as:
$$
Q^H_{\rm{ex}} = \frac{\sin^N\alpha}{V^N}  \int {\rm{exp}}\left\{{-\beta \sum 
\frac{\gamma_T}{2}
\left[(x_i-x_i^0)^2+(y_i-y_i^0)^2\right]}\right\}
\prod \rd x_i \rd y_i =  \frac{\sin^N\alpha}{V^N} \left(\frac{2\piup}{\beta \gamma_T}\right)^N.
$$
The corresponding free energy then is:
$$
F_H=-k T N \log \left(\frac{\sin\alpha}{V} \frac{2 \piup}{\beta \gamma_T}\right) = 
-k T N \log \left( \frac{2 \piup}{ V \beta \gamma_T}\right)
- k T N \log (\sin\alpha),
$$
where $k$ is the Boltzmann constant.
In this expression, we singled 
out
the second term that depends on~$\alpha$.
The first term will vanish in calculations of free energy
 differences using the same $\gamma_T$ but 
different~$\alpha$.
Thus, the harmonic free energy relevant for our calculations is 
just 
$- k T N \log (\sin\alpha) $.
The final formula for
the free energy becomes:
\begin{equation}
\Delta F=
- k T N \log (\sin\alpha)
+ {\displaystyle \int_0^1}\langle\Delta U\rangle_{\lambda} \rd \lambda.
\label{ecm1}
\end{equation}
For ellipses,
the methodology remains the same  except that now we need 
to add $N$ new degrees of freedom --- particle angles. The ideal part of the 
free energy will get new terms 
arising from new variables 
(comprising the inertia tensor of the
ellipses)
but these are independent of $\alpha$, so they can be neglected. In
the excess part, we 
will get additional integration over the angles with normalization 
constant $(2 \piup )^N$. The harmonic potential will now also apply to the angles:
$$U_{\rm H}={\displaystyle \sum_{i=1}^{N}}\left\{ \frac{\gamma_T}{2}\left[(x_i-x_i^0)^2+(y_i-y_i^0)^2\right]
+ \frac{\gamma_R}{2}(\phi_i-\phi_i^0)^2 \right\},
$$
where $\gamma_R$ is an additional spring constant and $\phi_i^0$ are the set of initial
particle angles.
The new partition function changes into:
$$
Q^H_{\rm{ex}} = 
 \frac{\sin^N\alpha}{V^N (2 \piup)^N} \left(\frac{2\piup}{\beta \gamma_T}\right)^N \left(\frac{2\piup}{\beta \gamma_R}\right)^{N/2}
$$
but the free energy part that depends on $\alpha$ remains the same: 
$
- k T N \log (\sin\alpha)$.

In the course of simulations, we discovered
that the integrand $\langle\Delta U\rangle_{\lambda}$ 
in formula~(\ref{ecm1}) converges very inefficiently for $\lambda=1$, which 
corresponds to the system driven by the
hard-core potential while its potential energy  is evaluated with the help of the harmonic potential.
The potential energy in this case experiences strong fluctuations that decay very slowly over time.
After some research, the slow convergence was tracked down to the movements of the simulation cell
as a whole. Since the random displacements of the particles during MC steps are uncorrelated, the center
of mass of the system experiences displacements from its initial position as well.
These displacements average out over time because of the law of large numbers but it may
take a long simulation time in order to see that. For non-zero $\lambda$'s, the collective movements
of all particles are not an issue because the harmonic potential suppresses large-scale deviations
from the initial coordinates. At the same time, the displacements of all particles as a whole
do not create a physically distinct state and thus should not affect the relative free energy. It,
therefore, makes sense to transition to the reference frame associated with the center of mass
of the system. The description then includes the coordinates of the center of mass $x_c$, $y_c$ and $N-1$ 
coordinates of the particles which characterize their mutual arrangement (conformations) 
(the coordinates of the remaining particle are expressed in terms
 of these  new
variables). Under periodic boundary conditions, the center of mass samples from the volume
$V/N$, where $V$ is the total volume of the system. This is the quantity that will appear in front
of the integral~(\ref{exess}) when the integration is carried out over $x_c$ and $y_c$. Importantly,
this volume per particle is the same for all $\alpha$, so the associated term of free energy 
will drop when the difference is taken.
Integration over the remaining $N-1$ coordinates yields:
$$Q^H_{\rm{ex}} = 
 \frac{\sin^{N-1}\alpha}{V^{N-1} (2 \piup)^N} 
\left(\frac{2\piup}{\beta \gamma_T}\right)^{N-1} \left(\frac{2\piup}{\beta \gamma_R}\right)^{N/2}$$
with the relevant $\alpha$-dependent term of free energy 
$- k T (N-1) \log (\sin\alpha)$. This leads to 
the final expression~(\ref{ecm})
for the free energy 
given in the section~\ref{sec:Methods}.
The function $\langle\Delta U\rangle_{\lambda}$ is evaluated in simulations carried out in the reference frame of the
center of mass. In practice, this was achieved by aligning the coordinates of the center
of mass at each MC step. We found that this can be accomplished by keeping track of two sets of
coordinates: one containing real coordinates and the other storing coordinates that 
are periodically imaged. 
This prevented discontinuous jumps of the center of mass position when particles were put back into the
simulation cell by the boundary conditions.
We found that formula~(\ref{ecm}) produces the same free energy difference as 
formula~(\ref{ecm1}) but 
at a much lower computational cost. For systems with $N>100$,  formula~(\ref{ecm1}) could not be
converged at all. We emphasize that formula~(\ref{ecm}) applies only when $\gamma_T$ and $\gamma_R$ are
taken to be the same for different $\alpha$'s. Additionally, they should be
sufficiently large, so that analytical integration in the partition function applies. In practice,
this can be achieved by gradually increasing $\gamma_T$ and $\gamma_R$ while monitoring
$\langle\Delta U\rangle_{\lambda}$ at $\lambda=0$. The point at which this function becomes equal
to the energy of the harmonic approximation $\frac{1}{2} k T N_F$ indicates that the spring constants
are strong enough. Here, $N_F$ is the number of degrees of freedom in the system. For the model
where all particles are allowed to rotate $N_F=2 N-2 + N=3 N -2$ and for the model with coupled
rotations $N_F=2 N-2 + 1=2N-1$.

Integration of $\langle\Delta U\rangle_{\lambda}$ was carried out numerically. 
To counter a very strong decline of  $\langle\Delta U\rangle_{\lambda}$ in the limit of $\lambda \to 1$,
the following change of variables was performed:
\begin{equation}
\left\{
\begin{array}{c}
\zeta =\frac{a_t}{b_t-1} \frac{1}{(1 + \alpha_t-\lambda)^{b_t-1} }, \\
\lambda = 1 + \alpha_t - (\frac{a_t}{b_t-1} \frac{1}{\zeta})^{{1}/{(b_t-1)}}. \\ 
\end{array}
\right.
\end{equation}
The integral in equation~(\ref{ecm}) was transformed accordingly:
\begin{equation}
\int_0^1{\langle\Delta U\rangle_{\lambda} } \rd \lambda = 
\int_{\zeta(0)}^{\zeta(1)}{\langle\Delta U\rangle_{\zeta} } 
\frac{1}{a_t}  \left( \frac{a_t}{b_t-1} \frac{1}{\zeta} \right) ^{{b_t}/{(b_t-1)}}
\rd \zeta =
\int_{\zeta(0)}^{\zeta(1)}{ F(\zeta) \rd \zeta },
\label{et}
\end{equation}
where parameters $a_t=1287.4$ and $b_t=0.44708$ were determined by fitting.
The constant $\alpha_t$ was set equal  $10^{-6}$.

Both $\langle\Delta U\rangle_{\zeta}$
and the derivative of this function $\frac{\rd \langle\Delta U\rangle_{\zeta}}{\rd \zeta}$ were used to compute the
integral. The latter can be extracted directly from simulations through the following
relationship: 
$$\frac{\rd \langle\Delta U\rangle_{\zeta}}{\rd \zeta}=-\beta ( \langle\Delta U^2\rangle_{\zeta} - \langle\Delta U\rangle_{\zeta}^2).$$
The derivative of the integrand in equation~(\ref{et}) can be found as
$$F'(\zeta)=\frac{\rd \langle\Delta U\rangle_{\zeta}}{\rd \zeta} \frac{1}{a_t} \left( \frac{a_t}{b_t-1} \frac{1}{\zeta} \right)
^{{b_t}/{(b_t-1)}} - \frac{b_t}{a_t^2}
  \left( \frac{a_t}{b_t-1} \frac{1}{\zeta} \right)^{{(2 b_t-1)}/{(b_t-1)}}
\langle\Delta U\rangle_{\zeta}.
$$

Numerical integration was carried out by 
the Euler--Maclaurin
method~\cite{phys}.
Initially, there were 15 non-overlapping segments considered with 
widths adjusted iteratively to achieve 
$\Delta \zeta_i = \zeta_{i+1}-\zeta_i \sim {1}/{|F(\zeta_i)|},\ i=0,14$. This allowed us
to make the numerical error almost uniform across the integration range.
Each segment was integrated using the 2-point formula 
$P_2 = 
h ({f_1}/{2}+{f_{2}}/{2})+{h^2}(p_1-p_2)/{12}$,
where 
$h=\zeta_2-\zeta_1$,
$f_1$ and $p_1$ are the 
values of the 
integrand and its first derivative 
at the first point of the segment and  $f_2$ and $p_2$ are the corresponding quantities
at the second point of the segment.
Each segment was then split evenly in two, yielding an intermediate point
$\zeta'=\frac{1}{2}(\zeta_1 + \zeta_2)$, and integration was repeated using the 3-point
formula $P_3= h ({f_1}/{2}+f'+{f_{2}}/{2})+{h^2}(p_1-p_2)/{12}$,
where $f'$ is the value of the function at $\zeta'$ and $h=\frac{1}{2} (\zeta_2-\zeta_1)$.
Since the Euler--Maclaurin formula is accurate up to $O(h^4)$, the two estimates, $I=P_2 + \alpha' h^4$ and
$I=P_3 + \frac{1}{16} \alpha' h^4$ can be combined in a 
 Romberg-style procedure to yield
a better approximation for the integral
$I={(16 P_3- P_2)}/{15}$,
which  is accurate up to terms $O(h^6)$.
The error contained in this estimate can be approximated by  
${(P_3-P_2)}/{15}$, which is the error of the 3-point integration formula.
If the error estimated in this way 
turned out to be higher than a pre-set target value, we 
continued to split the concerned segments until a desired accuracy was reached.
A total of 33 integration points were generated in this manner.
The grid was much denser for $\zeta$ points corresponding to $\lambda \sim 1$.
The estimated integration error is less than 0.1\%  in relative terms. For the free energy
difference, this translates into a 3\% numerical error, which is about twice as low as 
the statistical error resulting from incomplete sampling.

%\bibliographystyle{cmpj}
%\bibliography{library}

\ukrainianpart

\title{ До алгоритму проведення Монте Карло симуляцій в комірках моделювання з постійним об'ємом та змінною формою }
\author{А. Баумкетнер } 
\address{
	Інститут фізики конденсованих систем НАН України, вул. Свєнціцького, 1, 79011, Львів, Україна 
}

\makeukrtitle
\begin{abstract}
 В симуляціях кристалів певні властивості досліджуваної системи можуть залежати не тільки від
 об'єму комірки моделювання але й від її форми.
 В таких випадках бажано змінювати форму комірки в процесі симуляцій, оскільки 
 наперед вона може бути невідомою. 
 У цій роботі описано алгоритм який дозволяє це робити 
 з тої умови, щоб відтворити ключові параметри
 форми комірки, які спостерігаються в ансамблі при постійному тиску. 
 Алгоритм протестовано в симуляціях системи твердих еліпсів, яка може утворювати
 гратки різного типу. Показано, що використання запропонованого алгоритму призводить до
 доброго узгодження відносної вільної енергії різних типів граток з результатами отриманими
 незалежними методами.

\keywords{ твердий еліпс, Монте Карло, ансамбль при постійному об'ємі
}

\end{abstract}

\lastpage
\end{document}